\documentclass[11pt,a4paper]{article}
\pdfoutput=1
\usepackage{jcappub}
\usepackage{graphicx}
\usepackage{amsmath}   
\usepackage{amssymb}

\def\eq#1{{Eq.~(\ref{#1})}}

\title{Constraining the host galaxy halos of massive black holes from LISA event rates}

\author[a, b]{Hamsa Padmanabhan,}

\author[c]{Abraham Loeb}

\affiliation[a]{Canadian Institute for Theoretical Astrophysics \\
60 St. George Street, Toronto, ON M5S 3H8, Canada}
\affiliation[b]{D\'epartement de Physique Th\'eorique, Universite de Gen\`eve \\
24 quai Ernest-Ansermet, CH 1211 Gen\`eve 4, Switzerland}

\affiliation[c]{Astronomy department, Harvard University \\
60 Garden Street, Cambridge, MA 02138, USA}

\emailAdd{hamsa.padmanabhan@unige.ch}
\emailAdd{aloeb@cfa.harvard.edu}

\abstract{
The coalescence of massive black hole binaries (with masses $10^4 - 10^7 M_{\odot}$) leads to gravitational wave emission that is detectable out to high redshifts ($z \sim 20$) with the forthcoming LISA observatory. We combine the theoretically derived merger rates for dark matter haloes at various 
redshifts, with an empirically motivated prescription that connects the mass of a dark matter halo and that of its central black hole. Using the  expected constraints on the (chirp or reduced) masses of binary black holes, their mass ratios and redshift uncertainties, we forecast the measurement precision on the occupation fraction, normalization and slope of the black hole mass - halo mass relation at various redshifts, assuming a five-year LISA survey for three different confidence scenarios. We use the expected sizes of the LISA localization ellipses on the sky to estimate the number of electromagnetic counterparts to the gravitational wave sources which are detectable by future wide-field optical surveys, such as LSST. 
}

\begin{document}
\maketitle
\flushbottom

\section{Introduction}

Observations have now established that supermassive black holes inhabit the centres of most galaxies out to high redshifts \citep[e.g.,][]{ferrarese2000, magorrian1998, kormendy2013, tremaine2002, fan2001}.  In the standard hierarchical structure formation scenario, the assembly of galaxies takes place via the repeated coalescences of their host dark matter haloes. Thus, massive black hole binaries, formed from the merger of galaxies each containing a massive black hole, are expected to be ubiquitous throughout cosmic time \citep{begelman1980, haehnelt1994, mirosavljevic2001, wyithe2003, sesana2007, tanaka2009, kulkarni2012, barausse2012, antoine2016}. Direct observational evidence \citep{komossa2003} for supermassive black hole binaries has now been found both locally \citep[e.g.,][]{kharb2017} and at intermediate redshifts, $z \sim 0.2$ \citep{goulding2019}. The coalescence of  binaries in the mass range $(10^4 - 10^7 M_{\odot})$ leads to the emission of gravitational radiation 
at mHz frequencies, which is detectable by 
the Laser Interferometry Space Antenna (LISA) observatory \citep{amaro2017} and the proposed TianQin space-borne detector \citep{luo2016, wang2019, feng2019}.  Such 
coalescences are expected to occur more frequently at high redshifts, since the 
mergers of dark matter haloes are expected to be higher at early times \citep{volonteri2003}. LISA will be able to detect binary 
black hole mergers out to redshifts $z > 20$ with signal-to-noise ratios (SNRs) $\gtrsim 10$, and 
SNRs approaching $\sim 1000$ at low redshifts. The number of such events per year is estimated to be of the order of a few to a few thousand, though there is a large variation in the predictions of different models \citep[e.g.,][]{menou2001, haehnelt2003, jaffe2003,  wyithe2003, sesana2005, Lippai2008, barausse2020}.

Gravitational wave measurements from massive black hole binary mergers with the LISA observatory will allow to infer several properties, such as the binary members' masses (or, equivalently, the chirp mass and reduced mass of the system), spin vectors, rough sky location of the merger, and the source luminosity distance (which can be converted into a redshift for an assumed cosmology) at high precision 
\citep[e.g.,][]{cutler1998, hughes2002, vecchio2004, berti2005, arun2007, lang2011}. These measurements promise exciting new information constraining the seeds for the first supermassive black holes, their dynamical evolution and their relation to the observed luminosity function of quasars \citep{micic2007,menou2001, islam2004, koushiappas2006, sesana2007}. Subsets of the measured parameters are often highly correlated with each other, thus making it difficult to isolate a source from the entire population of coalescence events. It was, however, shown in Ref. \citep{lang2011} that including precessional effects due to the interaction of one black hole's spin with the gravitomagnetic fields from the other hole's spin, breaks the degeneracies among several parameters, thus greatly improving the accuracy in their measurement. In particular, masses are measurable by LISA to accuracies of $10^{-4} - 10^{-5}$, and luminosity distances to $0.2\% - 0.7\%$ at $z \sim 1$. Including the information in the spin precession also leads to an improvement in the localization of the sources on the sky to error ellipses with major axes of several tens of arcminutes, and minor axes a factor $2-4$ times smaller. 
 If an electromagnetic 
counterpart to the gravitational wave emission is found, the  LISA 
sources are also expected to act as `standard sirens', enabling a measurement of the expansion history 
of the universe \citep[e.g.,][]{chen2018} and
 uncovering valuable tests of General Relativity \citep[e.g.,][]{moorehellings2001, vecchio2004,hughes2002, tanaka2010, kocsis2006}.
 
 In this paper, we combine the theoretically derived merger rates for dark matter haloes at various 
redshifts, with empirical expressions connecting the mass of a dark matter halo and that of its central black hole \citep{wyithe2002}. Introducing an occupation fraction parameter, $f_{\rm bh}$ \citep{lippai2009}, that describes the fraction of host haloes that are expected to harbor central supermassive black holes, allows for an analytical computation of the merger rate of binary black holes as an explicit function of their masses, redshift and the parameters governing the black hole mass - halo mass relation. Using this merger rate in conjunction with the expected masses, ratios and redshift uncertainties \citep{lang2011} including the effects of precession, we forecast constraints  on the three parameters: (i) $f_{\rm bh}$, the occupation fraction of the black holes, 
(ii) 
$\gamma$, the power-law slope of the BH mass - halo mass relation when expressed in terms 
of the halo circular velocity $v_{\rm c,0}$, and (iii) $\epsilon_0$, the amplitude of this 
relation, from a five-year LISA survey for three different confidence 
scenarios. Finally, we use the expected 3D error ellipsoid of localization of the merger with a LISA survey  to place constraints on the expected number of electromagnetic counterparts to the gravitational wave sources, adopting 
LSST on the Vera Rubin Observatory as an example of a future wide-field optical survey for this purpose.

The paper is organized as follows. In Sec. \ref{sec:mergerrates}, we discuss the theoretical formalism involved in computing the merger rates of binary black holes  from that of their host dark matter haloes. We then, in Sec. \ref{sec:paramconstraints}, use the parameter constraints available in the literature \citep{lang2011} to place constraints on the free parameters of interest, given the expected level of uncertainty in the measurement of the remaining parameters. This requires a modification of the standard Fisher matrix procedure, which we describe in Sec. \ref{sec:setup}, before computing the constraints  in Sec. \ref{sec:computeconstraints} for three different confidence scenarios of LISA detection rates in Sec. \ref{sec:confscenarios} and Sec. \ref{sec:midscenario}. We constrain the expected number of electromagnetic counterpart galaxies to the binary black hole merger detectable in a future wide field survey like LSST  in Sec. \ref{sec:eleccounter}. Finally, we summarize our conclusions and discuss future prospects in Sec. \ref{sec:conclusions}.

\section{Merger rates of massive black holes}
\label{sec:mergerrates}
The number of gravitational wave sources detectable by LISA is a convolution of:
(i) the merger rate of the galaxies that contain black holes in the relevant  
mass range, and (ii) the occupation fraction of these galaxies, i.e. the fraction 
containing a black hole at their center. Since galaxies are known to reside in 
dark matter halos described by the hierarchical scenario of structure 
formation, the merger of halos is related to the coalescence rate of 
binary black holes.

We begin with the formalism for the merger rate of dark matter halos per unit 
redshift ($z$) and halo mass fraction ($\xi$), as formulated by \citep{fakhouri2010}:
\begin{eqnarray}
\frac{d n_{\rm halo}}{dz d\xi} &=& A \left(\frac{M}{10^{12} 
M_{\odot}}\right)^{\alpha} \xi^{\beta} 
\exp\left[\left(\frac{\xi}{\bar{\xi}}\right)^{\gamma_1}\right] (1+ z)^{\eta}
\label{halomerger_rate}
\end{eqnarray}
where $M$ is the mass of the primary halo, $\xi$ is the mass ratio of the two 
merging haloes, and the parameters 
have the values $\alpha = 0.133$, $\beta = -1.995$,
$\gamma_1 = 0.263$, $\eta = 0.0993$, $A = 0.0104$ and $\bar{\xi} = 9.72 \times 
10^{-3}$.

Combining the above rate with the abundance of haloes with masses between $M$ 
and $M + dM$, we can convert it into a merger rate per unit logarithmic halo 
mass, as:
\begin{eqnarray}
 \frac{d n_{\rm halo}}{d \log_{10} M dz d \xi}  = A \left(\frac{M}{10^{12} 
M_{\odot}}\right)^{\alpha} \xi^{\beta} \times 
\exp\left[\left(\frac{\xi}{\bar{\xi}}\right)^{\gamma_1}\right] (1+ z)^{\eta} 
\frac{dn_{\rm halo}}{d \log_{10} M}
\label{mergerratehaloes} \, ,
\end{eqnarray}
where  $dn_{\rm halo}/d \log_{10} M$ is the halo mass function, for which we adopt the Sheth Tormen 
form \citep{sheth2002}.

To predict the expected rate of binary black hole mergers, we combine the 
halo merger rate with the empirical 
relation connecting black hole and host halo mass \citep[e.g.,][]{wyithe2002}:
\begin{equation}
 M_{\rm BH} = M \epsilon_0 \left(\frac{M}{10^{12} M_{\odot}}\right)^{\gamma/3 - 
1} \left(\frac{\Delta_v \Omega_m h^2}{18 \pi^2}\right)^{\gamma/6} 
(1+z)^{\gamma/2} \, ,
\end{equation}
which is consistent with observations \citep[e.g.,][]{ferrarese2002} in the local 
universe and assumes a power-law scaling of the black hole mass with virial 
velocity, $M_{\rm BH} \propto v_{\rm c,0}^{\gamma}$. The above relation is also
consistent with the observed black hole - bulge mass relation 
\citep[Ref.][Sec. 6.10 and Eq. 10]{kormendy2013} coupled with the empirically derived stellar mass - halo 
mass relation \citep[e.g.,][]{behroozi2013}. With this, \eq{mergerratehaloes} can be 
recast as:
\begin{eqnarray}
&& \frac{d n_{\rm BH}}{d \log_{10} M_{\rm BH} dz dq}  = f_{\rm bh}^2 A_1 
\left(\frac{M_{\rm BH}}{10^{12} M_{\odot} K(z, \gamma, 
\epsilon_0)}\right)^{3\alpha/\gamma} \nonumber \\
&& \  \times q^{3/\gamma - 1 +3\beta/\gamma} 
(1+z)^{\eta} \exp\left[\left(\frac{q}{\bar{q}}\right)^{3\gamma_1/\gamma}\right] 
\frac{dn_{\rm halo}}{d \log_{10} M}
 \label{bhmergerrate}
\end{eqnarray}
where $q$ is the black hole mass ratio, related to $\xi$ as $q = \xi^{\gamma/3}$, $\bar{q}(\gamma) = 
\bar{\xi}^{\gamma/3}$ with $\bar{\xi} = 9.72 \times 10^{-3}$, $A_1 = (3/\gamma)^2 A$ and 
in which we have used the fact that $d \log_{10} M_{\rm BH} = (\gamma/3) d 
\log_{10} M$. We have also introduced the occupation fraction $f_{\rm bh}$, 
which measures the likelihood of merging 
haloes to contain black holes (which may in general, be a function of the halo 
mass, but is assumed to be constant here for simplicity, since the precise 
connection of the occupation fraction to host halo properties is currently 
unknown \citep{lippai2009}). This then relates $q$ to $\xi$ as $q = 
\xi^{\gamma/3}$. The function $K(z, \gamma, \epsilon_0)$ is defined as:
\begin{equation}
 K(z, \gamma, \epsilon_0) = \epsilon_0 \left(\frac{\Delta_v \Omega_m h^2}{18 
\pi^2}\right)^{\gamma/6} (1+z)^{\gamma/2}
\end{equation}
The merger rate of binary black holes is thus characterized by the three free 
parameters $f_{\rm bh}, \gamma, \epsilon_0$ (with the other two, $\eta$ and 
$\alpha$ being inherited 
from the underlying halo merger rate).

\section{Expected constraints}
\label{sec:paramconstraints}

The future LISA observatory will be able to detect several mergers of massive 
binary black holes at high redshifts through their gravitational wave 
emission in the milli-Hertz (mHz) frequency range. In this section, we 
illustrate the constraints that forthcoming LISA detections can be used to place 
on the properties of the black hole mass - halo mass relation and occupation fraction at 
high redshifts, described through the parameters $f_{\rm bh}, \gamma$ and  
$\epsilon_0$.

\subsection{Setting up the problem}
\label{sec:setup}

To begin with, we use \eq{bhmergerrate} to define the observed rate of GW events per comoving volume
per unit time \citep[e.g., Ref.][]{middleton2018}, as:
\begin{eqnarray}
 && \mathcal{R} (M_{\rm BH}, q, z; \epsilon_0, f_{\rm bh}, \gamma) =  \frac{d 
n_{\rm BH}}{d \log_{10} M_{\rm BH} dz dq} \frac{dz}{dt} \frac{1}{(1+z)} \nonumber \\
  \hskip-0.1in && =  H(z) f_{\rm bh}^2 A_1 (\gamma) \left(\frac{M_{\rm 
BH}}{10^{12} M_{\odot} K(z, \gamma, \epsilon_0)}\right)^{3\alpha/\gamma} \nonumber \\
&& \times q^{3/\gamma - 1 +3\beta/\gamma} 
(1+z)^{\eta} \exp\left[\left(\frac{q}{\bar{q} 
(\gamma)}\right)^{3\gamma_1/\gamma}\right] \frac{dn}{d \log_{10} M}   
\label{obsrate}
\end{eqnarray}
in which $H(z)$ is the Hubble parameter at redshift $z$, and we have made the 
parameter dependences explicit: $A_1(\gamma) = (3/\gamma)^2 A$ with $A = 0.0104$,  
$\bar{q}(\gamma) = 
\bar{\xi}^{\gamma/3}$ with $\bar{\xi} = 9.72 \times 10^{-3}$, and other 
constants have the values defined previously in Eqs. (\ref{halomerger_rate}) and 
(\ref{bhmergerrate}). Note that the observed rate differs from the intrinsic rate by the redshift dilation factor of $(1+z)$.

It is known \citep[see, for example, Ref.][]{lang2011} that LISA measurements from 
$10^4$ binary black hole mergers with masses $10^5 -  3 \times 10^6 M_{\odot}$ 
over $z \sim 1-5$ can constrain the individual (chirp or reduced) black hole 
masses $M_{\rm BH}$, redshifts $z$ and mass ratios $q$, with a relative 
precision ranging from 0.1\% - 10\% in various scenarios. Our objective is to 
use this information, together with the rate equation above, to measure how well 
a given detection scenario can constrain the three free parameters $\epsilon_0$, $f_{\rm bh}$ and 
$\gamma$. In its most generic form, the above problem can be expressed as a 
constraint equation:
\begin{equation}
 f(K_i; U_j) = 0 \, ,
\end{equation}
on a function $f$ of (i) `known' variables $K_i$,  $i = 1$ to $n$, all of which are 
determinable to a specified degree of accuracy, i.e. $\Delta K_i/K_i$ is known for all 
$i$, and (ii) `unknown' variables $U_j$, j = 1 to $m$, which are the parameters we 
wish to estimate the errors on. In our present case, $f$ is the difference 
between the event rate $\mathcal{R}$ and its fiducial value, the $K_i$'s are the 
set $\{M_{\rm BH}, q, z\}$, the $U_j$'s are the set $\{\epsilon_0, f_{\rm 
bh}, \gamma\}$, and $i = j = 3$. Given $f$ and $\Delta K_i/K_i$ for all $i$, we 
need to estimate $\Delta U_j/U_j$ for all $j$.

\subsection{Parameter constraints}
\label{sec:computeconstraints}

\begin{figure}
 \begin{center}
  \includegraphics[width = \columnwidth]{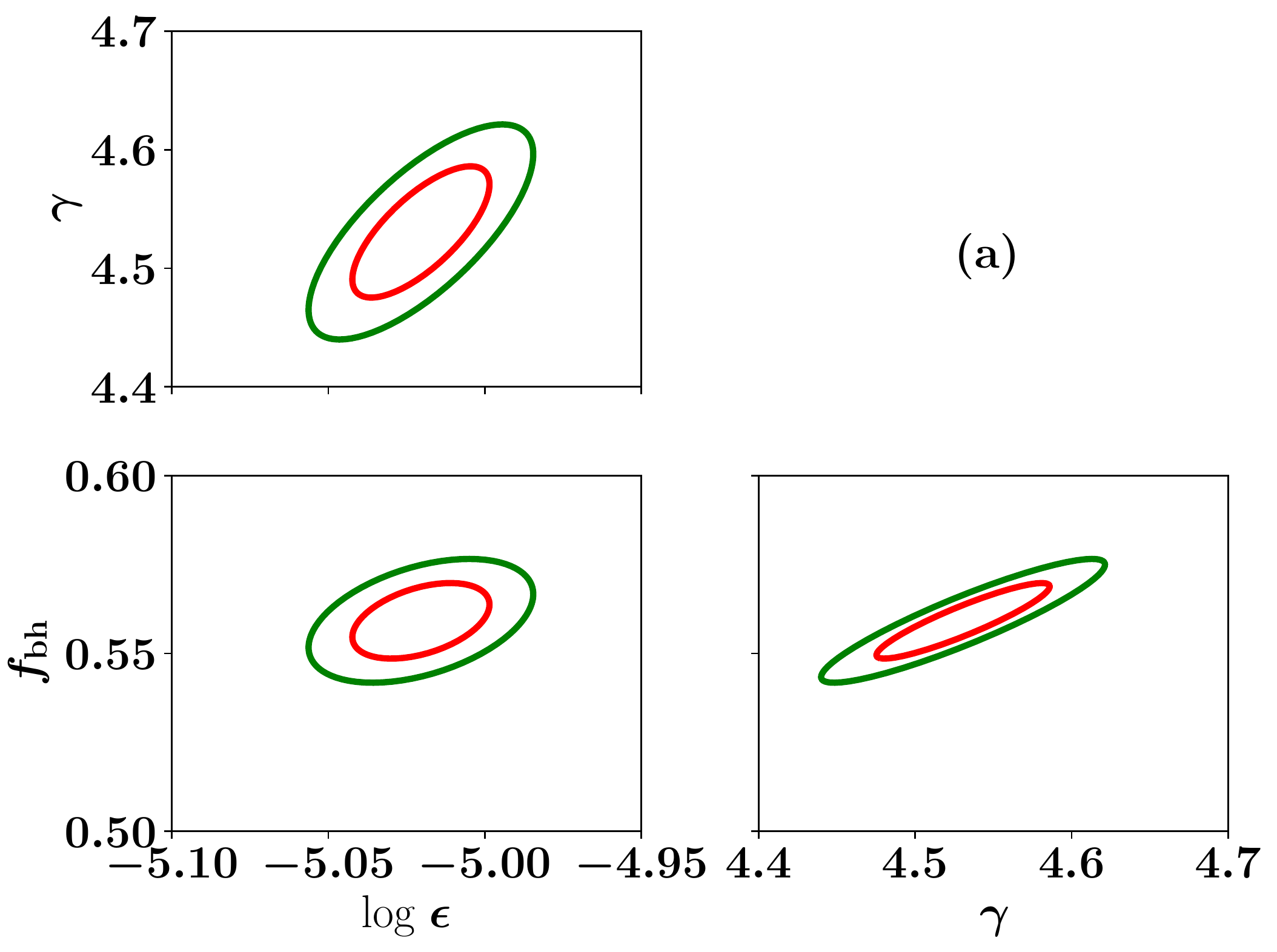} 
\includegraphics[width = \columnwidth]{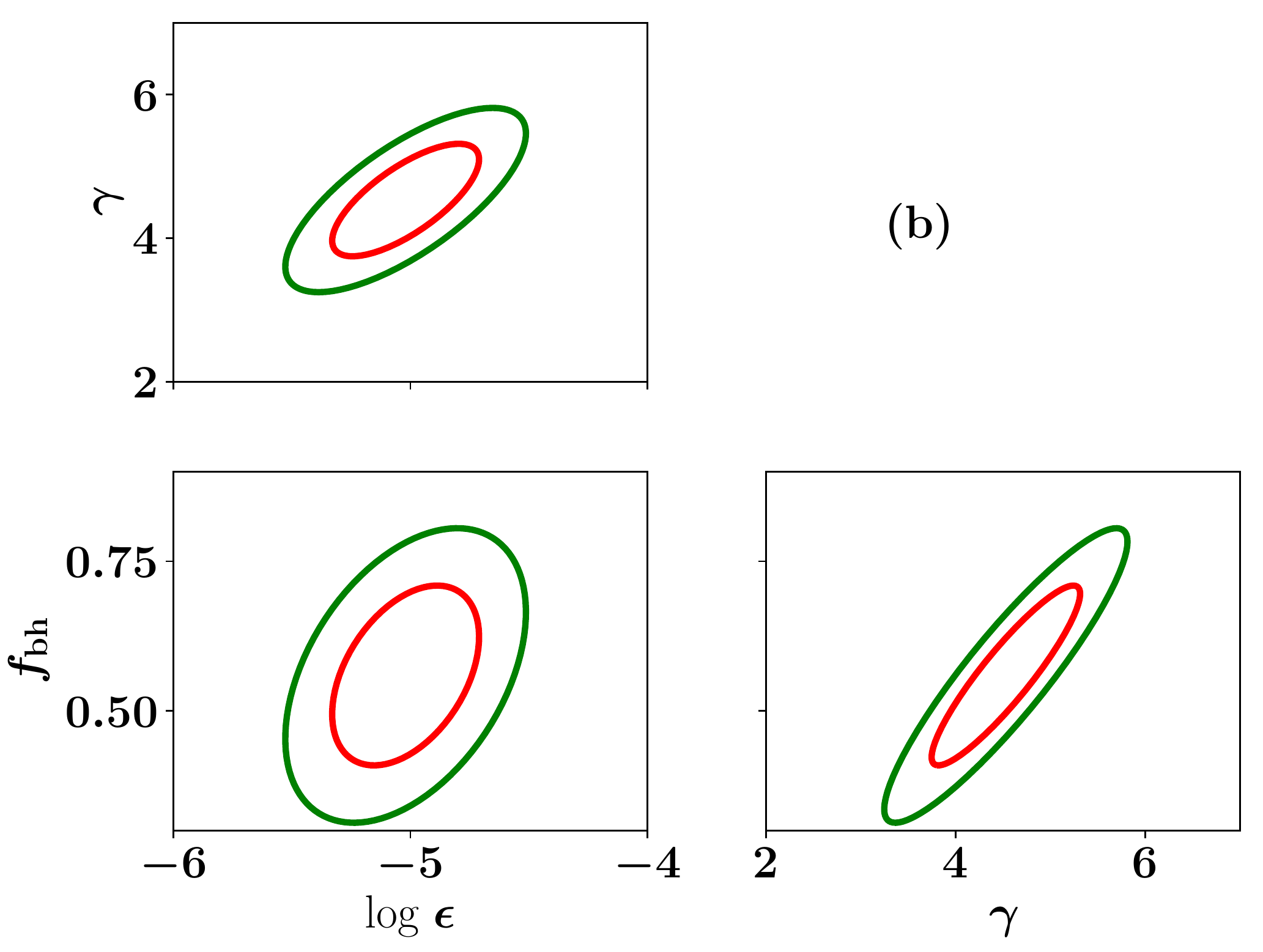}	
 \end{center}
 \caption{Extreme ends of the forecasting ability on the unknown parameters 
$\{\log \epsilon_0, \gamma, f_{\rm bh}\}$ with 
LISA. \textit{Panel (a)}: 1-$\sigma$ and 2-$\sigma$ confidence contours assuming the 
optimistic scenario for $\Delta \mathcal{R}$ and 400 LISA detections per year. 
\textit{Panel (b)}: Pessimistic scenario for $\Delta \mathcal{R}$ and 100 LISA 
detections per year.}
\label{fig:optpesscenarios}
\end{figure}

Towards handling the above 
(non-standard) situation, we refine the standard Fisher matrix formalism\footnote{Note that the approach described here is also useful in the generic scenario when: (i)  
one does not have a clearly defined likelihood function for the parameter 
constraints, and (ii) one is dealing with correlated parameters of which a 
subset are unknown, with the known ones being characterized by the probability 
distribution of their errors.} by summing the 
Fisher components at all known incidences of the `known' parameters.
We begin by using \eq{obsrate} to evaluate the observed event rate $\mathcal{R}$ 
in bins of $\log M_{\rm BH}$, $q$ and $z$, spaced over the relevant ranges in 
each of the known parameters $K_i$, where constraints are available 
\citep{lang2006} as
follows:\\
 $\log_{10} (M_{\rm BH}/M_{\odot}) = \{5,5.5, 6., 6.5\}$ \\
 $q = \{0.1, 0.3, 1.0\}$ \\
 $z = \{1,  3., 5.\}$ \\
This allows us to express the 
per-bin variation of $\mathcal{R}$ in each $\{\log_{10} M_{\rm BH}, q,z \}$ bin 
as:
 \begin{equation}
  \Delta \mathcal{R}_{\rm bin} = \frac{\partial \mathcal{R}}{\partial \log_{10} 
M_{\rm 
BH}} \Delta \log_{10} M_{\rm BH} + \frac{\partial \mathcal{R}}{\partial z} 
\Delta z + \frac{\partial \mathcal{R}}{\partial q} \Delta q \, ,
\label{deltarparameters}
 \end{equation}
 where the quantity on the RHS is evaluated in each bin.
 This step allows us to use the information available in the distributions 
of $\Delta K_i/K_i$, because we are counting all the incidences. This is 
equivalent to simulating the sample of individual values. The resultant standard 
deviation on $\mathcal{R}$ in each bin then becomes $\sigma_{\mathcal{R}_{\rm 
bin}} = \Delta \mathcal{R}_{\rm bin}/\sqrt{N_{\rm events}/N_{\rm bins}}$, where 
$N_{\rm events}$ is the number of events and $N_{\rm bins}$ is the number of 
bins. In practice, we may replace the $\Delta \mathcal{R}_{\rm bin}$ computed from \eq{deltarparameters} by an average value representing an assumed confidence scenario of uncertainties in the `known' parameters.

We can now use the Fisher formalism to forecast the expected uncertainties on 
the `unknowns' $\{\log \epsilon_0, \gamma, f_{\rm bh}\}$,\footnote{For ease of 
computation, we use $\log \epsilon_0$ instead of $\epsilon_0$ in the numerical 
results.} for varying numbers of LISA events in a 5-year survey. 
The fiducial values of these parameters, around which the errors are computed, 
are taken to be $U_{i, \rm fid} = \{-5.02, 
4.53, 0.56\}$ which are consistent with observations in nearby galaxies 
\citep{ferrarese2002, wyithe2002}. 
We calculate the $\{i,j\}$th element of the Fisher matrix $\mathcal{F}$, by 
summing over its contributions from each bin:

\begin{equation}
 \mathcal{F}_{ij} =  \sum_{\rm bin} \frac{1}{(\sigma_{\mathcal{R}_{\rm bin}})^2} 
\frac{\partial 
\mathcal{R}_{\rm bin}}{\partial U_i}\frac{\partial \mathcal{R}_{\rm bin}} 
{\partial U_j}
\label{fisher}
\end{equation}
where $U_i = \{\epsilon_0, f_{\rm bh}, \gamma\}$, and the derivatives $\partial 
\mathcal{R}_{\rm bin}/\partial U_i$ are computed in each bin $i$. \footnote{Note that this calculation makes the implict assumptions that the bins are independent since we neglect cross-correlations between the bins.}

\subsection{Confidence scenarios}
\label{sec:confscenarios}

The number of LISA detections predicted to take place every year is fairly 
uncertain \citep[e.g.,][]{katz2020, volonteri2020}.
Here, we consider three different confidence scenarios for the constraints on 
the 
parameters: (i) Optimistic, (ii) Pessimistic and (iii) Intermediate, which correspond to $(\Delta \mathcal{R}/\mathcal{R}) = 
0.001, 0.1$ and 0.01 
respectively for $10^4$ events, consistently with the expectations of 
Ref. \citep{lang2006, mangiagli2020}. Within each scenario, we further consider three 
different numbers 
of LISA events: 500, 1000 and 2000 respectively observed over a 5-year period
(corresponding to 100, 200 and 400 events per year) to compute 
$\sigma_{\mathcal{R}_{\rm bin}}$, and thus the elements of the Fisher matrix in \eq{fisher}, as 
detailed in the previous subsection.

From the Fisher matrix $\mathcal{F}$, we can now compute the standard deviations 
on each parameter, $p_i = \{K_i, U_i\}$, when the others are marginalized over, 
using the expression, $\sigma(p_i) = \sqrt{(\mathcal{F}^{-1})_{ii}}$. For the 
three `unknown' parameters $\{\log \epsilon_0, \gamma, f_{\rm bh}\}$, the values 
of $\sigma(U_i)/U_{i, \rm fid}$, with $U_{i, \rm fid} = \{-5.02, 4.53, 0.56\}$ 
are listed in Table \ref{table:optpesvalues}. The corresponding 1-$\sigma$  and 
2-$\sigma$ contours for the two extreme situations (pessimistic, with lowest 
number of events, and 
optimistic, with the highest number of events) are plotted in Fig. 
\ref{fig:optpesscenarios}.

\begin{figure}
 \begin{center}
  \includegraphics[width =\columnwidth]{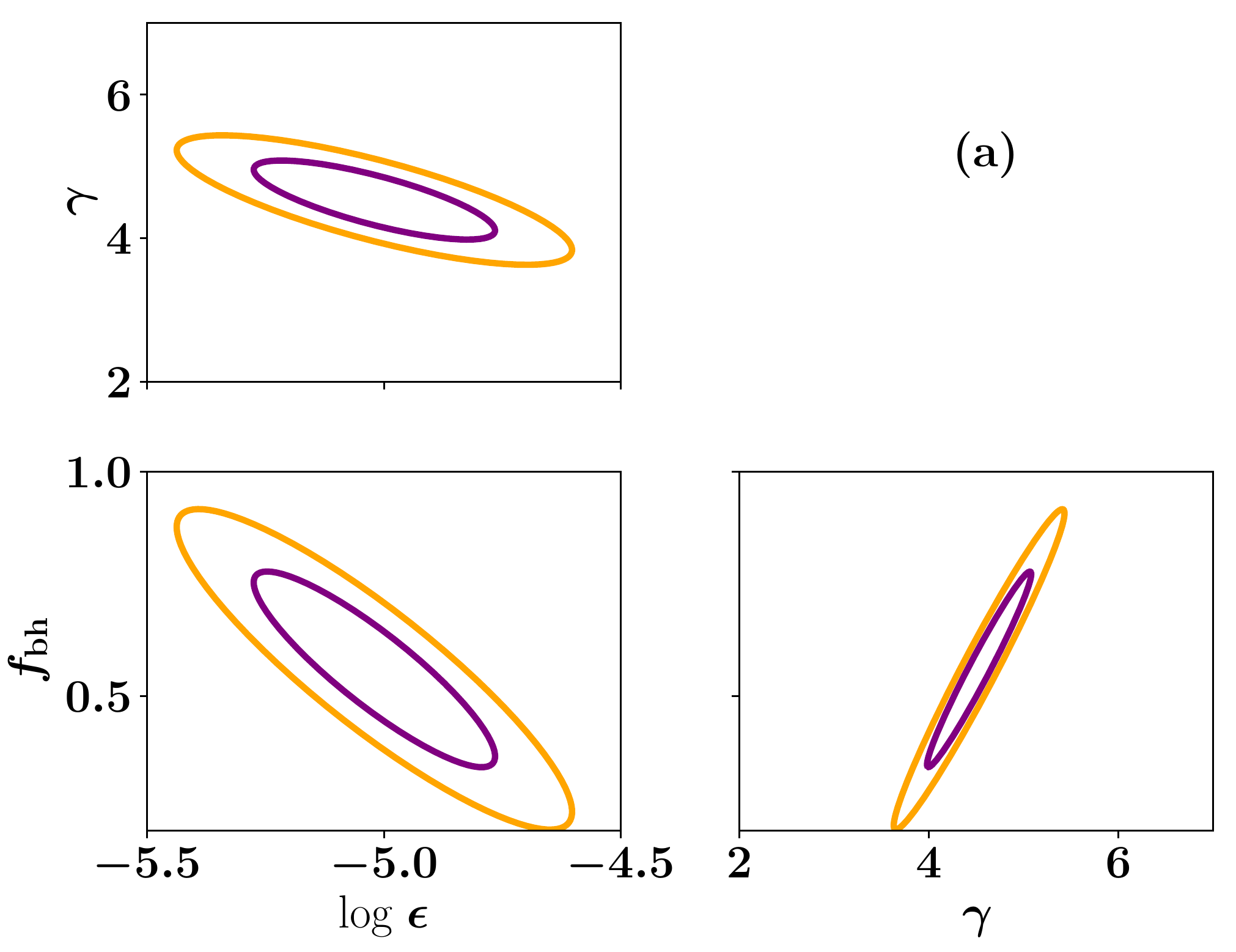} 
\includegraphics[width =\columnwidth]{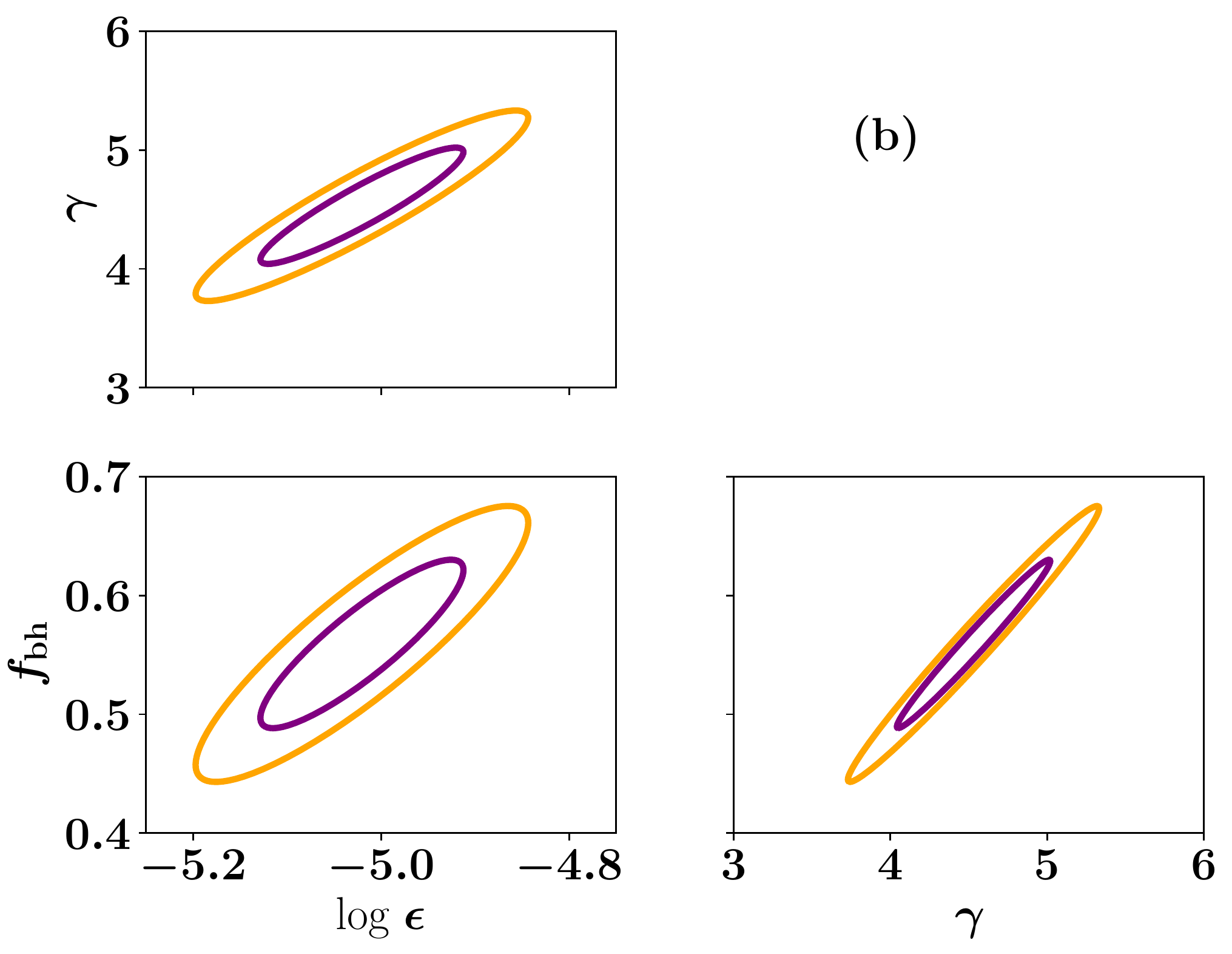}	
 \end{center}
 \caption{Intermediate or middle scenario for constraints on the unknown 
parameters, now focusing on two fixed values of redshift. \textit{Panel (a)}: 1-$\sigma$ 
and 2-$\sigma$ confidence contours assuming the intermediate number, i.e 200 
LISA detections per year, but focused on $z \sim 3$. \textit{Panel (b)}: Same 
as left panel, but focused on $z \sim 8$.}
\label{fig:redshift3and8}
\end{figure}

\subsection{Middle scenario}
\label{sec:midscenario}
We focus on an intermediate, or middle scenario to explore the constraints achievable for individual redshifts. 
In this scenario, the relative error is taken to be between the two cases 
considered above, at $\Delta \mathcal{R}/\mathcal{R} = 0.01$ (for $10^4$ events), and we fix the number of detections to 1000 (i.e. 200 per year observed over 5 years). 
Within this scenario, 
we consider two individual redshifts, one at $z \sim 3$ (which lies within the range in which constraints are available) and another at $z \sim 8$ (which lies outside this range), in order to illustrate the possible 
evolution of the constraining ability with respect to redshift.\footnote{The $z 
\sim 8$ case is to be considered very optimistic since the errors 
on the parameters are expected to be far worse by then \citep{lang2006}.} These are shown in Fig. \ref{fig:redshift3and8} and Table \ref{table:redshift3and8values}. 
Some of the constraints improve upon reaching higher
redshifts, $z \gg 1$, assuming that the uncertainties on the known parameters continue to 
hold.

\begin{table*}

 \begin{tabular}{ccccccccc}
  Pessimistic &$\sigma(U_i)/U_{i, \rm fid}$ & & & &  Optimistic &$\sigma(U_i)/U_{i, \rm fid}$ & & \\
 \hline
  No. of events & $\log \epsilon_0$ & $\gamma$ & $f_{\rm bh}$  & &  No. of 
events 
& $\log \epsilon_0$ & $\gamma$ & $f_{\rm bh}$ \\
  \hline \\
  500 & 0.017 & 0.022 & 0.046 & & 500 &  0.0017 & 0.0022 & 0.0046 \\
  1000 & 0.015 & 0.019 & 0.038 & & 1000 &  0.0015 & 0.0019 & 0.0038\\
  2000 & 0.012 & 0.016 & 0.032 & & 2000 &  0.001 & 0.002 & 0.003\\
  \hline
 \end{tabular}
 \caption{Expected relative errors, $\sigma(U_i)/U_i$, on the `unknown' 
parameters $\{\log \epsilon_0, \gamma, f_{\rm bh}\}$ around their fiducial 
values $U_{i, \rm fid} = \{-5.02, 4.53, 0.56\}$, for the pessimistic and 
optimistic scenarios considered in the main text. In each case, the forecasted 
parameter constraints assume a 5-year LISA survey with 100, 200 and 400 events 
per year.}
 \label{table:optpesvalues}
\end{table*}

\begin{table}
 \begin{center}
 \begin{tabular}{cccc}
   & & $\sigma(U_i)/U_{i, \rm fid}$ &  \\
 \hline
  Redshift & $\log \epsilon_0$ & $\gamma$ & $f_{\rm bh}$   \\
  \hline \\
  $z \sim 3$ & 0.006 & 0.007 & 0.016  \\
  $z \sim 8$ & 0.005 & 0.009 & 0.014 \\
  \hline
 \end{tabular}
 \end{center}
 \caption{Expected relative errors, $\sigma(U_i)/U_i$, on the `unknown' 
parameters $\{\log \epsilon_0, \gamma, f_{\rm bh}\}$ around their fiducial 
values $U_{i, \rm fid} = \{-5.02, 4.53, 0.56\}$, for the `middle scenario' at two individual redshifts, $z \sim 3$ and $z \sim 8$. The forecasts 
assume a 5-year LISA survey with 200 events 
per year.}
 \label{table:redshift3and8values}
\end{table}

\section{Localization and electromagnetic counterparts}
\label{sec:eleccounter}

Binary neutron star mergers detected in gravitational waves 
are expected to have an electromagnetic counterpart \citep[e.g,][]{li1998,
metzger2010, coulter2017, soares2017,soares2019}, which allows for the identification of a 
host galaxy.
It was shown that gravitational wave observations by LIGO-Virgo from the merger of two neutron stars have the potential to constrain 
the Hubble constant to within a few percent in five years, if a host galaxy is 
identified, either from a direct electromagnetic counterpart or from a 
statistical analysis of a  catalogue of potential host galaxies \citep[e.g.,][]{chen2018, abbott2017, gray2020, nair2018, nissanke2010, nissanke2013, di2018, mortlock2019, fishbach2019, holz2005, farr2019, feeney2019, dalal2006, singer2014}. Electromagnetic counterparts to
stellar-origin binary black holes have also been predicted in the literature \citep[e.g.,][]{orazio2018, loeb2016} via GRB afterglows. Another possible electromagnetic 
counterpart to the LIGO candidate event S190521g was reported recently by the 
Zwicky Transient Factory \citep[ZTF;][]{graham2020}.

Here, we explore the possibility of identifying the potential host galaxies of 
LISA-detected binary supermassive black holes, using  a catalogue of candidate 
electromagnetic counterparts detected by a future photometric survey, using 
 LSST on the Vera Rubin Observatory\footnote{www.lsst.org} as an example. We focus throughout on electromagnetic counterparts from the stellar light of the host galaxy (which is much longer lived than the transient counterparts arising from the emission from hot gas around the compact objects). If the 
electromagnetic counterpart can be unambigously identified, its sky position and 
redshift can be measured accurately. In the absence of a precise identification,  
we combine the expected uncertainties in the error ellipse parameters and 
$\Delta z/(1 + z)$  (from Ref. \citep{lang2011}) with the number of galaxies per unit sky 
area in the relevant range detectable by LSST, to estimate the total number of 
potential host galaxies needed to be searched for in order to identify the counterpart.

The properties of the detected sources are modelled using the LSST redshift 
selection function \citep{chang2013}, which is modified from the form in the 
LSST Science Book \citep{abell2009} for galaxies having $i$-band magnitudes 
$20.5 < i < 25.5$:
\begin{equation}
\phi(z) \propto z^{1.28} \exp\left(-\frac{z}{0.41}\right)^{0.97}
\end{equation} 
The corresponding galaxy surface number density is derived from the stellar mass correpsonding to each black hole mass bin. This is calculated using the results of Ref. \citep{behroozi2019} to assign stellar masses, $M_{*}$ to the black hole host dark matter haloes, which in turn are derived from the black hole mass - halo mass relation of Ref. \citep{wyithe2002}. The derived stellar masses are converted into $i$-band magnitudes using the $M_*/L_i- (g-i)$ relation of Ref. \citep{taylor2011}, assuming a typical $(g-i) = 1.5$ for LSST-detected spiral galaxies \citep[LSST Science Book, Ref.][Table 3.1]{abell2009}. The $K$-correction is added following the estimates\footnote{http://kcor.sai.msu.ru/} for $z \sim 0.5$, consistently with the SDSS findings from Fig. 6 of Ref. \citep{omill2011}, which also indicates  evidence that the $K$-correction flattens at higher redshifts.

Noting that the uncertainty on the $M_{\rm BH} - M_{*}$ is of the order of 0.29 dex \citep{kormendy2013} and that on the $(g-i) - L_i/M_{*}$ is 0.1 dex, the combined uncertainty on the i-magnitudes is $\sim 0.30$ dex. The error on the $M_{\rm BH}$ values from LISA is very small, of the order of sub-percent \citep{lang2011}. Using this range in the $i$-magnitudes as the $i$-bin widths, we can now estimate the number of galaxies per square arcmin within each $i$- and redshift bin, using the formula $N(<i) = 46\times 10^{0.31(i - 25)}$ arcmin$^{-2}$ from the LSST Science Book \citep[Ref.][Eq. 3.7]{abell2009} multiplied by the redshift selection function of LSST above.

\begin{figure}
 \begin{center}
  \includegraphics[scale=0.5]{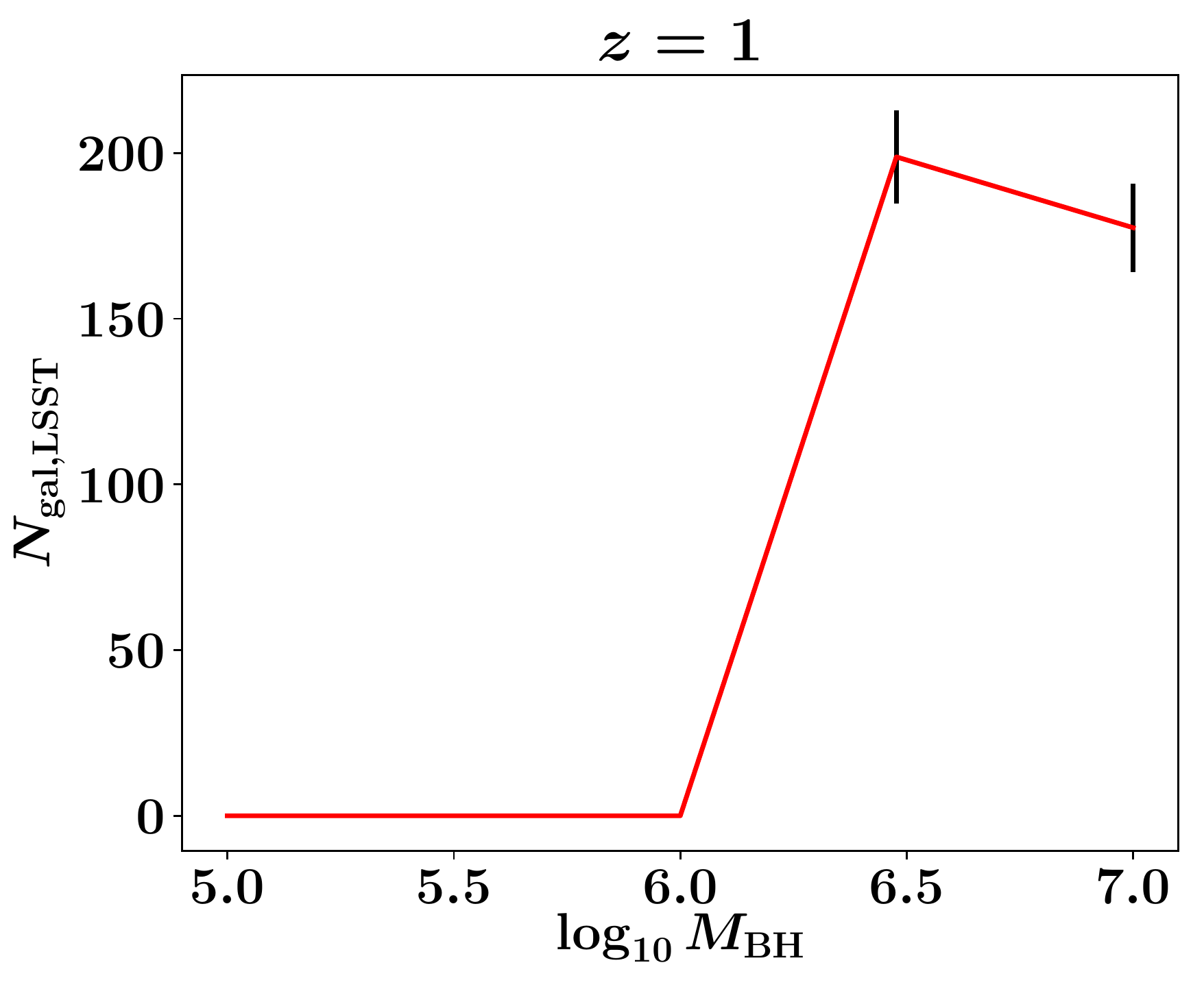}
 \end{center}
\caption{Expected number of host galaxies detectable by LSST at $z \sim 1$, as a 
function of primary black hole mass measured by LISA. The  median errors in the 
sky positions (ellipse major and minor axes $2a$ and $2b$ respectively) and redshift localizations of 
binary black holes with LISA \citep{lang2011} are combined with the expected 
number counts of galaxies in the same interval detectable by LSST. Note that detections are not possible in the first three mass bins since they correspond to $i$-band luminosities fainter than the LSST sensitivity limit (assumed here to be $i < 26.5$ for LSST-deep).}
\label{fig:lsstlisa}
\end{figure}

From the results of binary black hole merger analyses (e.g., 
Ref. \citep[][]{lang2011}, see Table IV), the median values of the major and minor 
ellipse axes ($2a$ and $2b$) for sources with black hole masses in the range 
$\{10^5, 10^7\}  M_{\odot}$
are expected to be in the range 13 arcmin to about 81 arcmin. Assuming a 
redshift localization of $\Delta z/z = 0.01$ around $z \sim 1$ (of the order of 
the uncertainty in $\Delta d_L/d_L$), \footnote{The relation between luminosity 
distance and redshift is nonlinear and dependent on the assumed cosmology - 
which can, in turn, be constrained, if an electromagnetic counterpart is found. 
For simplicity in the present analysis, we assume the same order of magnitude of 
the estimated relative errors in $\Delta d_L/d_L$ and $\Delta z/z$, noting that 
for a standard $\Lambda CDM$ cosmology consistent with the latest constraints 
\citep{planck, wmap7}, $\Delta d_L/d_L \sim 0.01$ corresponds to $\Delta z/z 
\sim 0.0081$ at $z \sim 1$.} we find the expected mean number of LSST sources in each error ellipsoid as 
a function of black hole mass, as shown in Fig. \ref{fig:lsstlisa}. The plot 
shows that LSST is expected to detect of order $\sim 100-200$ galaxies with
the black hole masses above $10^{6.5} M_{\odot}$.

Electromagnetic counterparts of supermassive binary black hole mergers in gas 
disks have been well studied in the literature 
\citep[e.g.,][]{bogdanovic2008, rossi2010, corrales2010}. It is possible that 
galaxies that are intrinsically fainter than the LSST limit above will enter the 
regime of detectability due to the bright flare caused by the binary black hole 
merger, which is dependent on the gas content and other properties of the host 
\citep{corrales2010, kocsis2008}, including the gas density profile and feedback effects. This 
in turn,  opens up the possibility of subsequent follow-up searches for the host 
galaxy with other instruments.

\section{Discussion}
\label{sec:conclusions}

In this paper, we have computed the rate of massive binary black hole mergers (of masses $10^4 - 10^7 M_{\odot})$ out to high redshifts ($z \gtrsim 5$) and connected it up to the number of events detectable by the forthcoming LISA observatory. Our theoretical framework assumes black holes to be associated with host dark matter haloes, with the black hole mass scaling as a power law with the halo circular velocity, as needed to match the observed luminosity function of quasars \citep{wyithe2002, volonteri2003}. The parameters (normalization, $\epsilon_0$ and slope, $\gamma$) of this empirically motivated relation \citep{ferrarese2002}, which matches the latest constraints in the local universe \citep{kormendy2013} are assumed to hold to high redshifts as well \citep{shields2003}. 

The mergers of massive black holes are assumed to follow those of their underlying host dark matter haloes. Halo mergers \citep{fakhouri2010} are assumed to lead to black hole coalescence without delay. We also neglected the halo merger timescale, assuming it to be much shorter than the Hubble time. This is a valid assumption if the black hole binaries do not have extreme mass ratios ($q \equiv M_1/M_2 < 20)$, \citep[e.g.,][]{tanaka2009, wyithe2003, kulkarni2012}, we additionally expect extreme mass ratio inspirals (EMRIs) to be depleted because of the long dynamical friction times for small sub-haloes in big haloes \citep{wetzel2010, garrison2017} and thus restrict to $q > 0.1$ in the calculations. We introduce an occupation fraction parameter, $f_{\rm bh}$, which measures the probability that the dark matter halo hosts a seed black hole. Theoretical models based on merger trees have shown that values of $f_{\rm bh} \gtrsim 0.1$ can accurately reproduce the evolution of the quasar luminosity function at redshifts $0 < z < 6$, as well as the mass function of remnant supermassive black holes at $z = 0$ \citep{lippai2009} including the effects of recoils from gravitational wave emission \citep{blecha2008} and triple systems \citep{kulkarni2012}.

Given the expected uncertainties in the measurement of black hole masses and ratios at various redshifts (the `known' parameters), one can use the LISA detection rate to place constraints on the remaining `unknown' parameters, viz. the occupation fraction ($f_{\rm bh}$), normalization ($\epsilon_0$) and slope ($\gamma$) of the black hole - halo mass relation. To evaluate the prospects for this goal, we modified the standard Fisher matrix approach and accounted for the available constraints on the subset of `known' parameters in three different confidence scenarios, each assumed to have 100, 200 or 400 event detections per year, for a survey of a 5 year duration. In so doing, we have found that the occupation fraction of black holes ($f_{\rm bh}$) and the parameters governing the black hole mass to halo mass evolution ($\epsilon_0$ and $\gamma$) can be constrained to percent or sub-percent levels of accuracy around $z \sim 1-5$, depending on the scenario under consideration. If the uncertainties on the measured source parameters are assumed to hold to higher 
redshifts ($z \sim 8$), then the parameter constraints become tighter.

We have also explored the possibility of detecting the electromagnetic 
counterpart from the host galaxy stellar light associated with the massive or supermassive binary black hole merger, 
using future wide-field photometric surveys, such as  LSST. We assumed that the electromagnetic follow up occurs in the post-merger period, once the black holes have reached coalescence. Given the expected range of sensitivity of LSST, and an assumed conversion between the black hole  masses, their host galaxy masses and the corresponding $i$-band magnitudes, we expect roughly 100-200 electromagnetic counterparts to fall within the expected LSST senstitivity range for binaries with masses above $M_{\rm BH} \sim 10^{6.5} M_{\odot}$ at $z \sim 1$. These figures are comparable to the estimates derived by Ref. \citep{kocsis2008}, who address a different problem: that of monitoring the LISA sources with LSST in the 2-3 weeks \textit{preceding} the merger.  Another difference is that the analysis of Ref. \citep{kocsis2008} considers the  gas accretion emission from the merger whereas we focus on stellar light. However, after accounting for a localization cut corresponding to the LISA error ellipse, and imposing photometric redshift and luminosity bounds (which makes their assumptions fairly comparable to those in our present analysis), their derived number of counterpart candidates reaches $N_{\rm counterpart} \sim 1-1000$, consistently with the present findings.

A caveat to the error estimates is the assumed Gaussian approximation to the likelihood function, which is the basis for the direct computation of the Fisher information matrix. This approximation is almost certain to underestimate the true uncertainty since it misses the possible long tail of the likelihood function, and is known to break down at low number counts \citep[e.g.,][]{lang2011}. Thus, the results presented here should be taken as optimistic. {{There is also the possibility of multiple supermassive black hole binaries residing in the same host halo \citep[e.g.,][]{hoffman2007, kulkarni2012} with corresponding gravitational-wave emission signatures \citep[e.g.,][]{wyithe2003, sesana2004, amaro2010}. The probability of multiple black holes in the nuclei of galaxies increases with increasing host halo mass and redshift.  Simulations \citep{kulkarni2012} show that only about 30\% of galaxies with haloes of masses $10^{11} M_{\odot}$ at $z \sim 6$ contain more than two supermassive black holes at redshifts $2 < z < 6$, while lower mass galaxies rarely host more than two supermassive black holes at any point in their assembly history.  Numerical simulations of triple black hole systems have been shown to produce distinct signatures in the gravitational wave spectrum \citep{amaro2010}, though their detectability with LISA relies on the development of adequate analysis techniques to extract the signal amidst the large confusion noise. The recoil associated with the gravitational wave emission \citep[e.g.,][]{blecha2008, oleary2009} can lead to escaping supermassive black holes.  About 10 percent \citep{kulkarni2012} of such black holes are ejected at velocities $> 2000$ km s$^{-1}$ and expected to spend a few Gyr in the outskirts of the halo. It would be interesting to consider the above effects in simulation forecasts for LISA.}} A more detailed analysis would also include the timescales for mergers and address the dependence of the parameter constraints on the estimated time before the merger \citep[e.g.,][]{mangiagli2020}, which we leave to future work.

\section*{Acknowledgements}
We thank the referee for their helpful comments. HP acknowledges support from the Swiss National Science Foundation through Ambizione Grant PZ00P2\_179934. The work of AL was partially supported by the Black Hole Initiative at Harvard 
University, which is funded by grants from the JTF and GBMF.

\def\aj{AJ}                   
\def\araa{ARA\&A}             
\def\apj{ApJ}                 
\def\apjl{ApJ}                
\def\apjs{ApJS}               
\def\ao{Appl.Optics}          
\def\apss{Ap\&SS}             
\def\aap{A\&A}                
\def\aapr{A\&A~Rev.}          
\def\aaps{A\&AS}              
\def\azh{AZh}                 
\def\baas{BAAS}
\def\jcap{JCAP}
\def\jrasc{JRASC}             
\def\memras{MmRAS}
\def\na{New Astronomy}
\def\nat{Nature}
\def\mnras{MNRAS}             
\def\pra{Phys.Rev.A}          
\def\prb{Phys.Rev.B}          
\def\prc{Phys.Rev.C}          
\def\prd{Phys.Rev.D}          
\def\prl{Phys.Rev.Lett}       
\def\pasp{PASP}               
\def\pasj{PASJ}
\def\physrep{Phys. Repts.}
\def\qjras{QJRAS}             
\def\skytel{S\&T}             
\def\solphys{Solar~Phys.}     
\def\sovast{Soviet~Ast.}      
\def\ssr{Space~Sci.Rev.}      
\def\zap{ZAp}                 
\let\astap=\aap
\let\apjlett=\apjl
\let\apjsupp=\apjs

\bibliography{mybib}{}

\providecommand{\href}[2]{#2}\begingroup\raggedright\begin{thebibliography}{10}

\bibitem{ferrarese2000}
L.~{Ferrarese} and D.~{Merritt}, \emph{{A Fundamental Relation between
  Supermassive Black Holes and Their Host Galaxies}},
  \href{https://doi.org/10.1086/312838}{\emph{\apjl} {\bfseries 539} (2000) L9}
  [\href{https://arxiv.org/abs/astro-ph/0006053}{{\ttfamily
  astro-ph/0006053}}].

\bibitem{magorrian1998}
J.~{Magorrian}, S.~{Tremaine}, D.~{Richstone}, R.~{Bender}, G.~{Bower},
  A.~{Dressler} et~al., \emph{{The Demography of Massive Dark Objects in Galaxy
  Centers}}, \href{https://doi.org/10.1086/300353}{\emph{\aj} {\bfseries 115}
  (1998) 2285} [\href{https://arxiv.org/abs/astro-ph/9708072}{{\ttfamily
  astro-ph/9708072}}].

\bibitem{kormendy2013}
J.~{Kormendy} and L.~C. {Ho}, \emph{{Coevolution (Or Not) of Supermassive Black
  Holes and Host Galaxies}},
  \href{https://doi.org/10.1146/annurev-astro-082708-101811}{\emph{\araa}
  {\bfseries 51} (2013) 511} [\href{https://arxiv.org/abs/1304.7762}{{\ttfamily
  1304.7762}}].

\bibitem{tremaine2002}
S.~{Tremaine}, K.~{Gebhardt}, R.~{Bender}, G.~{Bower}, A.~{Dressler}, S.~M.
  {Faber} et~al., \emph{{The Slope of the Black Hole Mass versus Velocity
  Dispersion Correlation}}, \href{https://doi.org/10.1086/341002}{\emph{\apj}
  {\bfseries 574} (2002) 740}
  [\href{https://arxiv.org/abs/astro-ph/0203468}{{\ttfamily
  astro-ph/0203468}}].

\bibitem{fan2001}
X.~{Fan}, V.~K. {Narayanan}, R.~H. {Lupton}, M.~A. {Strauss}, G.~R. {Knapp},
  R.~H. {Becker} et~al., \emph{{A Survey of z\&gt;5.8 Quasars in the Sloan
  Digital Sky Survey. I. Discovery of Three New Quasars and the Spatial Density
  of Luminous Quasars at z\raisebox{-0.5ex}\textasciitilde6}},
  \href{https://doi.org/10.1086/324111}{\emph{\aj} {\bfseries 122} (2001) 2833}
  [\href{https://arxiv.org/abs/astro-ph/0108063}{{\ttfamily
  astro-ph/0108063}}].

\bibitem{begelman1980}
M.~C. {Begelman}, R.~D. {Blandford} and M.~J. {Rees}, \emph{{Massive black hole
  binaries in active galactic nuclei}},
  \href{https://doi.org/10.1038/287307a0}{\emph{\nat} {\bfseries 287} (1980)
  307}.

\bibitem{haehnelt1994}
M.~G. {Haehnelt}, \emph{{Low-Frequency Gravitational Waves from Supermassive
  Black-Holes}}, \href{https://doi.org/10.1093/mnras/269.1.199}{\emph{\mnras}
  {\bfseries 269} (1994) 199}
  [\href{https://arxiv.org/abs/astro-ph/9405032}{{\ttfamily
  astro-ph/9405032}}].

\bibitem{mirosavljevic2001}
M.~{Milosavljevi{\'c}} and D.~{Merritt}, \emph{{Formation of Galactic Nuclei}},
  \href{https://doi.org/10.1086/323830}{\emph{\apj} {\bfseries 563} (2001) 34}
  [\href{https://arxiv.org/abs/astro-ph/0103350}{{\ttfamily
  astro-ph/0103350}}].

\bibitem{wyithe2003}
J.~S.~B. {Wyithe} and A.~{Loeb}, \emph{{Low-Frequency Gravitational Waves from
  Massive Black Hole Binaries: Predictions for LISA and Pulsar Timing Arrays}},
  \href{https://doi.org/10.1086/375187}{\emph{\apj} {\bfseries 590} (2003) 691}
  [\href{https://arxiv.org/abs/astro-ph/0211556}{{\ttfamily
  astro-ph/0211556}}].

\bibitem{sesana2007}
A.~{Sesana}, M.~{Volonteri} and F.~{Haardt}, \emph{{The imprint of massive
  black hole formation models on the LISA data stream}},
  \href{https://doi.org/10.1111/j.1365-2966.2007.11734.x}{\emph{\mnras}
  {\bfseries 377} (2007) 1711}
  [\href{https://arxiv.org/abs/astro-ph/0701556}{{\ttfamily
  astro-ph/0701556}}].

\bibitem{tanaka2009}
T.~{Tanaka} and Z.~{Haiman}, \emph{{The Assembly of Supermassive Black Holes at
  High Redshifts}},
  \href{https://doi.org/10.1088/0004-637X/696/2/1798}{\emph{\apj} {\bfseries
  696} (2009) 1798} [\href{https://arxiv.org/abs/0807.4702}{{\ttfamily
  0807.4702}}].

\bibitem{kulkarni2012}
G.~{Kulkarni} and A.~{Loeb}, \emph{{Formation of galactic nuclei with multiple
  supermassive black holes at high redshifts}},
  \href{https://doi.org/10.1111/j.1365-2966.2012.20699.x}{\emph{\mnras}
  {\bfseries 422} (2012) 1306}
  [\href{https://arxiv.org/abs/1107.0517}{{\ttfamily 1107.0517}}].

\bibitem{barausse2012}
E.~{Barausse}, \emph{{The evolution of massive black holes and their spins in
  their galactic hosts}},
  \href{https://doi.org/10.1111/j.1365-2966.2012.21057.x}{\emph{\mnras}
  {\bfseries 423} (2012) 2533}
  [\href{https://arxiv.org/abs/1201.5888}{{\ttfamily 1201.5888}}].

\bibitem{antoine2016}
A.~{Klein}, E.~{Barausse}, A.~{Sesana}, A.~{Petiteau}, E.~{Berti}, S.~{Babak}
  et~al., \emph{{Science with the space-based interferometer eLISA:
  Supermassive black hole binaries}},
  \href{https://doi.org/10.1103/PhysRevD.93.024003}{\emph{\prd} {\bfseries 93}
  (2016) 024003} [\href{https://arxiv.org/abs/1511.05581}{{\ttfamily
  1511.05581}}].

\bibitem{komossa2003}
S.~{Komossa}, \emph{{Observational evidence for supermassive black hole
  binaries}},  in \emph{The Astrophysics of Gravitational Wave Sources}, J.~M.
  {Centrella}, ed., vol.~686 of \emph{American Institute of Physics Conference
  Series}, pp.~161--174, Oct., 2003,
  \href{https://doi.org/10.1063/1.1629428}{DOI}
  [\href{https://arxiv.org/abs/astro-ph/0306439}{{\ttfamily
  astro-ph/0306439}}].

\bibitem{kharb2017}
P.~{Kharb}, D.~V. {Lal} and D.~{Merritt}, \emph{{A candidate sub-parsec binary
  black hole in the Seyfert galaxy NGC 7674}},
  \href{https://doi.org/10.1038/s41550-017-0256-4}{\emph{Nature Astronomy}
  {\bfseries 1} (2017) 727} [\href{https://arxiv.org/abs/1709.06258}{{\ttfamily
  1709.06258}}].

\bibitem{goulding2019}
A.~D. {Goulding}, K.~{Pardo}, J.~E. {Greene}, C.~M.~F. {Mingarelli},
  K.~{Nyland} and M.~A. {Strauss}, \emph{{Discovery of a Close-separation
  Binary Quasar at the Heart of a z 0.2 Merging Galaxy and Its Implications for
  Low-frequency Gravitational Waves}},
  \href{https://doi.org/10.3847/2041-8213/ab2a14}{\emph{\apjl} {\bfseries 879}
  (2019) L21} [\href{https://arxiv.org/abs/1907.03757}{{\ttfamily
  1907.03757}}].

\bibitem{amaro2017}
P.~{Amaro-Seoane}, H.~{Audley}, S.~{Babak}, J.~{Baker}, E.~{Barausse},
  P.~{Bender} et~al., \emph{{Laser Interferometer Space Antenna}}, {\emph{arXiv
  e-prints} (2017) arXiv:1702.00786}
  [\href{https://arxiv.org/abs/1702.00786}{{\ttfamily 1702.00786}}].

\bibitem{luo2016}
J.~{Luo}, L.-S. {Chen}, H.-Z. {Duan}, Y.-G. {Gong}, S.~{Hu}, J.~{Ji} et~al.,
  \emph{{TianQin: a space-borne gravitational wave detector}},
  \href{https://doi.org/10.1088/0264-9381/33/3/035010}{\emph{Classical and
  Quantum Gravity} {\bfseries 33} (2016) 035010}
  [\href{https://arxiv.org/abs/1512.02076}{{\ttfamily 1512.02076}}].

\bibitem{wang2019}
H.-T. {Wang}, Z.~{Jiang}, A.~{Sesana}, E.~{Barausse}, S.-J. {Huang}, Y.-F.
  {Wang} et~al., \emph{{Science with the TianQin observatory: Preliminary
  results on massive black hole binaries}},
  \href{https://doi.org/10.1103/PhysRevD.100.043003}{\emph{\prd} {\bfseries
  100} (2019) 043003} [\href{https://arxiv.org/abs/1902.04423}{{\ttfamily
  1902.04423}}].

\bibitem{feng2019}
W.-F. {Feng}, H.-T. {Wang}, X.-C. {Hu}, Y.-M. {Hu} and Y.~{Wang},
  \emph{{Preliminary study on parameter estimation accuracy of supermassive
  black hole binary inspirals for TianQin}},
  \href{https://doi.org/10.1103/PhysRevD.99.123002}{\emph{\prd} {\bfseries 99}
  (2019) 123002} [\href{https://arxiv.org/abs/1901.02159}{{\ttfamily
  1901.02159}}].

\bibitem{volonteri2003}
M.~{Volonteri}, F.~{Haardt} and P.~{Madau}, \emph{{The Assembly and Merging
  History of Supermassive Black Holes in Hierarchical Models of Galaxy
  Formation}}, \href{https://doi.org/10.1086/344675}{\emph{\apj} {\bfseries
  582} (2003) 559} [\href{https://arxiv.org/abs/astro-ph/0207276}{{\ttfamily
  astro-ph/0207276}}].

\bibitem{menou2001}
K.~{Menou}, Z.~{Haiman} and V.~K. {Narayanan}, \emph{{The Merger History of
  Supermassive Black Holes in Galaxies}},
  \href{https://doi.org/10.1086/322310}{\emph{\apj} {\bfseries 558} (2001) 535}
  [\href{https://arxiv.org/abs/astro-ph/0101196}{{\ttfamily
  astro-ph/0101196}}].

\bibitem{haehnelt2003}
M.~G. {Haehnelt}, \emph{{Hierarchical build-up of galactic bulges and the
  merging rate of supermassive binary black holes}},
  \href{https://doi.org/10.1088/0264-9381/20/10/304}{\emph{Classical and
  Quantum Gravity} {\bfseries 20} (2003) S31}
  [\href{https://arxiv.org/abs/astro-ph/0307379}{{\ttfamily
  astro-ph/0307379}}].

\bibitem{jaffe2003}
A.~H. {Jaffe} and D.~C. {Backer}, \emph{{Gravitational Waves Probe the
  Coalescence Rate of Massive Black Hole Binaries}},
  \href{https://doi.org/10.1086/345443}{\emph{\apj} {\bfseries 583} (2003) 616}
  [\href{https://arxiv.org/abs/astro-ph/0210148}{{\ttfamily
  astro-ph/0210148}}].

\bibitem{sesana2005}
A.~{Sesana}, F.~{Haardt}, P.~{Madau} and M.~{Volonteri}, \emph{{The
  Gravitational Wave Signal from Massive Black Hole Binaries and Its
  Contribution to the LISA Data Stream}},
  \href{https://doi.org/10.1086/428492}{\emph{\apj} {\bfseries 623} (2005) 23}
  [\href{https://arxiv.org/abs/astro-ph/0409255}{{\ttfamily
  astro-ph/0409255}}].

\bibitem{Lippai2008}
Z.~{Lippai}, Z.~{Frei} and Z.~{Haiman}, \emph{{Prompt Shocks in the Gas Disk
  around a Recoiling Supermassive Black Hole Binary}},
  \href{https://doi.org/10.1086/587034}{\emph{\apjl} {\bfseries 676} (2008) L5}
  [\href{https://arxiv.org/abs/0801.0739}{{\ttfamily 0801.0739}}].

\bibitem{barausse2020}
E.~{Barausse}, I.~{Dvorkin}, M.~{Tremmel}, M.~{Volonteri} and M.~{Bonetti},
  \emph{{Massive black hole merger rates: the effect of kpc separation
  wandering and supernova feedback}}, {\emph{arXiv e-prints} (2020)
  arXiv:2006.03065} [\href{https://arxiv.org/abs/2006.03065}{{\ttfamily
  2006.03065}}].

\bibitem{cutler1998}
C.~Cutler, \emph{Angular resolution of the lisa gravitational wave detector},
  \href{https://doi.org/10.1103/PhysRevD.57.7089}{\emph{Phys. Rev. D}
  {\bfseries 57} (1998) 7089}.

\bibitem{hughes2002}
S.~A. Hughes, \emph{{Untangling the merger history of massive black holes with
  LISA}}, \href{https://doi.org/10.1046/j.1365-8711.2002.05247.x}{\emph{Monthly
  Notices of the Royal Astronomical Society} {\bfseries 331} (2002) 805}.

\bibitem{vecchio2004}
A.~Vecchio, \emph{Lisa observations of rapidly spinning massive black hole
  binary systems},
  \href{https://doi.org/10.1103/PhysRevD.70.042001}{\emph{Phys. Rev. D}
  {\bfseries 70} (2004) 042001}.

\bibitem{berti2005}
E.~{Berti}, A.~{Buonanno} and C.~M. {Will}, \emph{{Testing general relativity
  and probing the merger history of massive black holes with LISA}},
  \href{https://doi.org/10.1088/0264-9381/22/18/S08}{\emph{Classical and
  Quantum Gravity} {\bfseries 22} (2005) S943}
  [\href{https://arxiv.org/abs/gr-qc/0504017}{{\ttfamily gr-qc/0504017}}].

\bibitem{arun2007}
K.~G. Arun, B.~R. Iyer, B.~S. Sathyaprakash, S.~Sinha and C.~Van Den~Broeck,
  \emph{Higher signal harmonics, lisa's angular resolution, and dark energy},
  \href{https://doi.org/10.1103/PhysRevD.76.104016}{\emph{Phys. Rev. D}
  {\bfseries 76} (2007) 104016}.

\bibitem{lang2011}
R.~N. {Lang}, S.~A. {Hughes} and N.~J. {Cornish}, \emph{{Measuring parameters
  of massive black hole binaries with partially aligned spins}},
  \href{https://doi.org/10.1103/PhysRevD.84.022002}{\emph{\prd} {\bfseries 84}
  (2011) 022002} [\href{https://arxiv.org/abs/1101.3591}{{\ttfamily
  1101.3591}}].

\bibitem{micic2007}
M.~{Micic}, K.~{Holley-Bockelmann}, S.~{Sigurdsson} and T.~{Abel},
  \emph{{Supermassive black hole growth and merger rates from cosmological
  N-body simulations}},
  \href{https://doi.org/10.1111/j.1365-2966.2007.12162.x}{\emph{\mnras}
  {\bfseries 380} (2007) 1533}
  [\href{https://arxiv.org/abs/astro-ph/0703540}{{\ttfamily
  astro-ph/0703540}}].

\bibitem{islam2004}
R.~R. {Islam}, J.~E. {Taylor} and J.~{Silk}, \emph{{Massive black hole remnants
  of the first stars - III. Observational signatures from the past}},
  \href{https://doi.org/10.1111/j.1365-2966.2004.08227.x}{\emph{\mnras}
  {\bfseries 354} (2004) 629}
  [\href{https://arxiv.org/abs/astro-ph/0309559}{{\ttfamily
  astro-ph/0309559}}].

\bibitem{koushiappas2006}
S.~M. {Koushiappas} and A.~R. {Zentner}, \emph{{Testing Models of Supermassive
  Black Hole Seed Formation through Gravity Waves}},
  \href{https://doi.org/10.1086/499325}{\emph{\apj} {\bfseries 639} (2006) 7}
  [\href{https://arxiv.org/abs/astro-ph/0503511}{{\ttfamily
  astro-ph/0503511}}].

\bibitem{chen2018}
H.-Y. {Chen}, M.~{Fishbach} and D.~E. {Holz}, \emph{{A two per cent Hubble
  constant measurement from standard sirens within five years}},
  \href{https://doi.org/10.1038/s41586-018-0606-0}{\emph{\nat} {\bfseries 562}
  (2018) 545} [\href{https://arxiv.org/abs/1712.06531}{{\ttfamily
  1712.06531}}].

\bibitem{moorehellings2001}
T.~A. Moore and R.~W. Hellings, \emph{Angular resolution of space-based
  gravitational wave detectors},
  \href{https://doi.org/10.1103/PhysRevD.65.062001}{\emph{Phys. Rev. D}
  {\bfseries 65} (2002) 062001}.

\bibitem{tanaka2010}
T.~{Tanaka} and K.~{Menou}, \emph{{Time-dependent Models for the Afterglows of
  Massive Black Hole Mergers}},
  \href{https://doi.org/10.1088/0004-637X/714/1/404}{\emph{\apj} {\bfseries
  714} (2010) 404} [\href{https://arxiv.org/abs/0912.2054}{{\ttfamily
  0912.2054}}].

\bibitem{kocsis2006}
B.~{Kocsis}, Z.~{Frei}, Z.~{Haiman} and K.~{Menou}, \emph{{Finding the
  Electromagnetic Counterparts of Cosmological Standard Sirens}},
  \href{https://doi.org/10.1086/498236}{\emph{\apj} {\bfseries 637} (2006) 27}
  [\href{https://arxiv.org/abs/astro-ph/0505394}{{\ttfamily
  astro-ph/0505394}}].

\bibitem{wyithe2002}
J.~S.~B. {Wyithe} and A.~{Loeb}, \emph{{A Physical Model for the Luminosity
  Function of High-Redshift Quasars}},
  \href{https://doi.org/10.1086/344249}{\emph{\apj} {\bfseries 581} (2002) 886}
  [\href{https://arxiv.org/abs/astro-ph/0206154}{{\ttfamily
  astro-ph/0206154}}].

\bibitem{lippai2009}
Z.~{Lippai}, Z.~{Frei} and Z.~{Haiman}, \emph{{On the Occupation Fraction of
  Seed Black Holes in High-redshift Dark Matter Halos}},
  \href{https://doi.org/10.1088/0004-637X/701/1/360}{\emph{\apj} {\bfseries
  701} (2009) 360}.

\bibitem{fakhouri2010}
O.~{Fakhouri}, C.-P. {Ma} and M.~{Boylan-Kolchin}, \emph{{The merger rates and
  mass assembly histories of dark matter haloes in the two Millennium
  simulations}},
  \href{https://doi.org/10.1111/j.1365-2966.2010.16859.x}{\emph{\mnras}
  {\bfseries 406} (2010) 2267}
  [\href{https://arxiv.org/abs/1001.2304}{{\ttfamily 1001.2304}}].

\bibitem{sheth2002}
R.~K. {Sheth} and G.~{Tormen}, \emph{{An excursion set model of hierarchical
  clustering: ellipsoidal collapse and the moving barrier}},
  \href{https://doi.org/10.1046/j.1365-8711.2002.04950.x}{\emph{\mnras}
  {\bfseries 329} (2002) 61}
  [\href{https://arxiv.org/abs/astro-ph/0105113}{{\ttfamily
  astro-ph/0105113}}].

\bibitem{ferrarese2002}
L.~{Ferrarese}, \emph{{Beyond the Bulge: A Fundamental Relation between
  Supermassive Black Holes and Dark Matter Halos}},
  \href{https://doi.org/10.1086/342308}{\emph{\apj} {\bfseries 578} (2002) 90}
  [\href{https://arxiv.org/abs/astro-ph/0203469}{{\ttfamily
  astro-ph/0203469}}].

\bibitem{behroozi2013}
P.~S. {Behroozi}, R.~H. {Wechsler} and C.~{Conroy}, \emph{{The Average Star
  Formation Histories of Galaxies in Dark Matter Halos from z = 0-8}},
  \href{https://doi.org/10.1088/0004-637X/770/1/57}{\emph{\apj} {\bfseries 770}
  (2013) 57} [\href{https://arxiv.org/abs/1207.6105}{{\ttfamily 1207.6105}}].

\bibitem{middleton2018}
H.~{Middleton}, S.~{Chen}, W.~{Del Pozzo}, A.~{Sesana} and A.~{Vecchio},
  \emph{{No tension between assembly models of super massive black hole
  binaries and pulsar observations}},
  \href{https://doi.org/10.1038/s41467-018-02916-7}{\emph{Nature
  Communications} {\bfseries 9} (2018) 573}
  [\href{https://arxiv.org/abs/1707.00623}{{\ttfamily 1707.00623}}].

\bibitem{lang2006}
R.~N. {Lang} and S.~A. {Hughes}, \emph{{Measuring coalescing massive binary
  black holes with gravitational waves: The impact of spin-induced
  precession}}, \href{https://doi.org/10.1103/PhysRevD.74.122001}{\emph{\prd}
  {\bfseries 74} (2006) 122001}
  [\href{https://arxiv.org/abs/gr-qc/0608062}{{\ttfamily gr-qc/0608062}}].

\bibitem{katz2020}
M.~L. {Katz}, L.~Z. {Kelley}, F.~{Dosopoulou}, S.~{Berry}, L.~{Blecha} and
  S.~L. {Larson}, \emph{{Probing massive black hole binary populations with
  LISA}}, \href{https://doi.org/10.1093/mnras/stz3102}{\emph{\mnras} {\bfseries
  491} (2020) 2301} [\href{https://arxiv.org/abs/1908.05779}{{\ttfamily
  1908.05779}}].

\bibitem{volonteri2020}
M.~{Volonteri}, H.~{Pfister}, R.~S. {Beckman}, Y.~{Dubois}, M.~{Colpi}, C.~J.
  {Conselice} et~al., \emph{{Black hole mergers from dwarf to massive galaxies
  with the NewHorizon and Horizon-AGN simulations}}, {\emph{arXiv e-prints}
  (2020) arXiv:2005.04902} [\href{https://arxiv.org/abs/2005.04902}{{\ttfamily
  2005.04902}}].

\bibitem{mangiagli2020}
A.~{Mangiagli}, A.~{Klein}, M.~{Bonetti}, M.~L. {Katz}, A.~{Sesana},
  M.~{Volonteri} et~al., \emph{{On the inspiral of coalescing massive black
  hole binaries with LISA in the era of Multi-Messenger Astrophysics}},
  {\emph{arXiv e-prints} (2020) arXiv:2006.12513}
  [\href{https://arxiv.org/abs/2006.12513}{{\ttfamily 2006.12513}}].

\bibitem{li1998}
L.-X. {Li} and B.~{Paczy{\'n}ski}, \emph{{Transient Events from Neutron Star
  Mergers}}, \href{https://doi.org/10.1086/311680}{\emph{\apjl} {\bfseries 507}
  (1998) L59} [\href{https://arxiv.org/abs/astro-ph/9807272}{{\ttfamily
  astro-ph/9807272}}].

\bibitem{metzger2010}
B.~D. {Metzger}, G.~{Mart{\'\i}nez-Pinedo}, S.~{Darbha}, E.~{Quataert},
  A.~{Arcones}, D.~{Kasen} et~al., \emph{{Electromagnetic counterparts of
  compact object mergers powered by the radioactive decay of r-process
  nuclei}},
  \href{https://doi.org/10.1111/j.1365-2966.2010.16864.x}{\emph{\mnras}
  {\bfseries 406} (2010) 2650}
  [\href{https://arxiv.org/abs/1001.5029}{{\ttfamily 1001.5029}}].

\bibitem{coulter2017}
D.~A. {Coulter}, R.~J. {Foley}, C.~D. {Kilpatrick}, M.~R. {Drout}, A.~L.
  {Piro}, B.~J. {Shappee} et~al., \emph{{Swope Supernova Survey 2017a (SSS17a),
  the optical counterpart to a gravitational wave source}},
  \href{https://doi.org/10.1126/science.aap9811}{\emph{Science} {\bfseries 358}
  (2017) 1556} [\href{https://arxiv.org/abs/1710.05452}{{\ttfamily
  1710.05452}}].

\bibitem{soares2017}
M.~{Soares-Santos}, D.~E. {Holz}, J.~{Annis}, R.~{Chornock}, K.~{Herner},
  E.~{Berger} et~al., \emph{{The Electromagnetic Counterpart of the Binary
  Neutron Star Merger LIGO/Virgo GW170817. I. Discovery of the Optical
  Counterpart Using the Dark Energy Camera}},
  \href{https://doi.org/10.3847/2041-8213/aa9059}{\emph{\apjl} {\bfseries 848}
  (2017) L16} [\href{https://arxiv.org/abs/1710.05459}{{\ttfamily
  1710.05459}}].

\bibitem{soares2019}
M.~{Soares-Santos}, {et al.}, {LIGO Scientific Collaboration} and {Virgo
  Collaboration}, \emph{{First Measurement of the Hubble Constant from a Dark
  Standard Siren using the Dark Energy Survey Galaxies and the LIGO/Virgo
  Binary-Black-hole Merger GW170814}},
  \href{https://doi.org/10.3847/2041-8213/ab14f1}{\emph{\apjl} {\bfseries 876}
  (2019) L7} [\href{https://arxiv.org/abs/1901.01540}{{\ttfamily 1901.01540}}].

\bibitem{abbott2017}
B.~P. {Abbott} and {et al.}, \emph{{A gravitational-wave standard siren
  measurement of the Hubble constant}},
  \href{https://doi.org/10.1038/nature24471}{\emph{\nat} {\bfseries 551} (2017)
  85} [\href{https://arxiv.org/abs/1710.05835}{{\ttfamily 1710.05835}}].

\bibitem{gray2020}
R.~{Gray}, I.~M. {Hernandez}, H.~{Qi}, A.~{Sur}, P.~R. {Brady}, H.-Y. {Chen}
  et~al., \emph{{Cosmological inference using gravitational wave standard
  sirens: A mock data analysis}},
  \href{https://doi.org/10.1103/PhysRevD.101.122001}{\emph{\prd} {\bfseries
  101} (2020) 122001} [\href{https://arxiv.org/abs/1908.06050}{{\ttfamily
  1908.06050}}].

\bibitem{nair2018}
R.~{Nair}, S.~{Bose} and T.~D. {Saini}, \emph{{Measuring the Hubble constant:
  Gravitational wave observations meet galaxy clustering}},
  \href{https://doi.org/10.1103/PhysRevD.98.023502}{\emph{\prd} {\bfseries 98}
  (2018) 023502} [\href{https://arxiv.org/abs/1804.06085}{{\ttfamily
  1804.06085}}].

\bibitem{nissanke2010}
S.~{Nissanke}, D.~E. {Holz}, S.~A. {Hughes}, N.~{Dalal} and J.~L. {Sievers},
  \emph{{Exploring Short Gamma-ray Bursts as Gravitational-wave Standard
  Sirens}}, \href{https://doi.org/10.1088/0004-637X/725/1/496}{\emph{\apj}
  {\bfseries 725} (2010) 496}
  [\href{https://arxiv.org/abs/0904.1017}{{\ttfamily 0904.1017}}].

\bibitem{nissanke2013}
S.~{Nissanke}, D.~E. {Holz}, N.~{Dalal}, S.~A. {Hughes}, J.~L. {Sievers} and
  C.~M. {Hirata}, \emph{{Determining the Hubble constant from gravitational
  wave observations of merging compact binaries}}, {\emph{arXiv e-prints}
  (2013) arXiv:1307.2638} [\href{https://arxiv.org/abs/1307.2638}{{\ttfamily
  1307.2638}}].

\bibitem{di2018}
E.~{Di Valentino}, D.~E. {Holz}, A.~r. {Melchiorri} and F.~{Renzi},
  \emph{{Cosmological impact of future constraints on H$_{0}$ from
  gravitational-wave standard sirens}},
  \href{https://doi.org/10.1103/PhysRevD.98.083523}{\emph{\prd} {\bfseries 98}
  (2018) 083523} [\href{https://arxiv.org/abs/1806.07463}{{\ttfamily
  1806.07463}}].

\bibitem{mortlock2019}
D.~J. Mortlock, S.~M. Feeney, H.~V. Peiris, A.~R. Williamson and S.~M.
  Nissanke, \emph{Unbiased hubble constant estimation from binary neutron star
  mergers}, \href{https://doi.org/10.1103/PhysRevD.100.103523}{\emph{Phys. Rev.
  D} {\bfseries 100} (2019) 103523}.

\bibitem{fishbach2019}
M.~{Fishbach}, R.~{Gray}, I.~{Maga{\~n}a Hernandez}, H.~{Qi}, A.~{Sur},
  F.~{Acernese} et~al., \emph{{A Standard Siren Measurement of the Hubble
  Constant from GW170817 without the Electromagnetic Counterpart}},
  \href{https://doi.org/10.3847/2041-8213/aaf96e}{\emph{\apjl} {\bfseries 871}
  (2019) L13} [\href{https://arxiv.org/abs/1807.05667}{{\ttfamily
  1807.05667}}].

\bibitem{holz2005}
D.~E. {Holz} and S.~A. {Hughes}, \emph{{Using Gravitational-Wave Standard
  Sirens}}, \href{https://doi.org/10.1086/431341}{\emph{\apj} {\bfseries 629}
  (2005) 15} [\href{https://arxiv.org/abs/astro-ph/0504616}{{\ttfamily
  astro-ph/0504616}}].

\bibitem{farr2019}
W.~M. {Farr}, M.~{Fishbach}, J.~{Ye} and D.~E. {Holz}, \emph{{A Future
  Percent-level Measurement of the Hubble Expansion at Redshift 0.8 with
  Advanced LIGO}}, \href{https://doi.org/10.3847/2041-8213/ab4284}{\emph{\apjl}
  {\bfseries 883} (2019) L42}
  [\href{https://arxiv.org/abs/1908.09084}{{\ttfamily 1908.09084}}].

\bibitem{feeney2019}
S.~M. {Feeney}, H.~V. {Peiris}, A.~R. {Williamson}, S.~M. {Nissanke}, D.~J.
  {Mortlock}, J.~{Alsing} et~al., \emph{{Prospects for Resolving the Hubble
  Constant Tension with Standard Sirens}},
  \href{https://doi.org/10.1103/PhysRevLett.122.061105}{\emph{\prl} {\bfseries
  122} (2019) 061105} [\href{https://arxiv.org/abs/1802.03404}{{\ttfamily
  1802.03404}}].

\bibitem{dalal2006}
N.~{Dalal}, D.~E. {Holz}, S.~A. {Hughes} and B.~{Jain}, \emph{{Short GRB and
  binary black hole standard sirens as a probe of dark energy}},
  \href{https://doi.org/10.1103/PhysRevD.74.063006}{\emph{\prd} {\bfseries 74}
  (2006) 063006} [\href{https://arxiv.org/abs/astro-ph/0601275}{{\ttfamily
  astro-ph/0601275}}].

\bibitem{singer2014}
L.~P. {Singer}, L.~R. {Price}, B.~{Farr}, A.~L. {Urban}, C.~{Pankow},
  S.~{Vitale} et~al., \emph{{The First Two Years of Electromagnetic Follow-up
  with Advanced LIGO and Virgo}},
  \href{https://doi.org/10.1088/0004-637X/795/2/105}{\emph{\apj} {\bfseries
  795} (2014) 105} [\href{https://arxiv.org/abs/1404.5623}{{\ttfamily
  1404.5623}}].

\bibitem{orazio2018}
D.~J. {D'Orazio} and A.~{Loeb}, \emph{{Single progenitor model for GW150914 and
  GW170104}}, \href{https://doi.org/10.1103/PhysRevD.97.083008}{\emph{\prd}
  {\bfseries 97} (2018) 083008}
  [\href{https://arxiv.org/abs/1706.04211}{{\ttfamily 1706.04211}}].

\bibitem{loeb2016}
A.~{Loeb}, \emph{{Electromagnetic Counterparts to Black Hole Mergers Detected
  by LIGO}}, \href{https://doi.org/10.3847/2041-8205/819/2/L21}{\emph{\apjl}
  {\bfseries 819} (2016) L21}
  [\href{https://arxiv.org/abs/1602.04735}{{\ttfamily 1602.04735}}].

\bibitem{graham2020}
M.~J. {Graham}, K.~E.~S. {Ford}, B.~{McKernan}, N.~P. {Ross}, D.~{Stern},
  K.~{Burdge} et~al., \emph{{Candidate Electromagnetic Counterpart to the
  Binary Black Hole Merger Gravitational Wave Event S190521g}}, {\emph{arXiv
  e-prints} (2020) arXiv:2006.14122}
  [\href{https://arxiv.org/abs/2006.14122}{{\ttfamily 2006.14122}}].

\bibitem{chang2013}
C.~{Chang}, M.~{Jarvis}, B.~{Jain}, S.~M. {Kahn}, D.~{Kirkby}, A.~{Connolly}
  et~al., \emph{{The effective number density of galaxies for weak lensing
  measurements in the LSST project}},
  \href{https://doi.org/10.1093/mnras/stt1156}{\emph{\mnras} {\bfseries 434}
  (2013) 2121} [\href{https://arxiv.org/abs/1305.0793}{{\ttfamily 1305.0793}}].

\bibitem{abell2009}
{LSST Science Collaboration}, P.~A. {Abell}, J.~{Allison}, S.~F. {Anderson},
  J.~R. {Andrew}, J.~R.~P. {Angel} et~al., \emph{{LSST Science Book, Version
  2.0}}, {\emph{arXiv e-prints} (2009) arXiv:0912.0201}
  [\href{https://arxiv.org/abs/0912.0201}{{\ttfamily 0912.0201}}].

\bibitem{behroozi2019}
P.~{Behroozi}, R.~H. {Wechsler}, A.~P. {Hearin} and C.~{Conroy},
  \emph{{UNIVERSEMACHINE: The correlation between galaxy growth and dark matter
  halo assembly from z = 0-10}},
  \href{https://doi.org/10.1093/mnras/stz1182}{\emph{\mnras} {\bfseries 488}
  (2019) 3143} [\href{https://arxiv.org/abs/1806.07893}{{\ttfamily
  1806.07893}}].

\bibitem{taylor2011}
E.~N. {Taylor}, A.~M. {Hopkins}, I.~K. {Baldry}, M.~J.~I. {Brown}, S.~P.
  {Driver}, L.~S. {Kelvin} et~al., \emph{{Galaxy And Mass Assembly (GAMA):
  stellar mass estimates}},
  \href{https://doi.org/10.1111/j.1365-2966.2011.19536.x}{\emph{\mnras}
  {\bfseries 418} (2011) 1587}
  [\href{https://arxiv.org/abs/1108.0635}{{\ttfamily 1108.0635}}].

\bibitem{omill2011}
A.~L. {O'Mill}, F.~{Duplancic}, D.~{Garc{\'\i}a Lambas} and J.~{Sodr{\'e}},
  Laerte, \emph{{Photometric redshifts and k-corrections for the Sloan Digital
  Sky Survey Data Release 7}},
  \href{https://doi.org/10.1111/j.1365-2966.2011.18222.x}{\emph{\mnras}
  {\bfseries 413} (2011) 1395}
  [\href{https://arxiv.org/abs/1012.3752}{{\ttfamily 1012.3752}}].

\bibitem{planck}
{Planck Collaboration}, P.~A.~R. {Ade}, N.~{Aghanim}, C.~{Armitage-Caplan},
  M.~{Arnaud}, M.~{Ashdown} et~al., \emph{{Planck 2013 results. XVI.
  Cosmological parameters}}, {\emph{arXiv:1303.5076} (2013) }
  [\href{https://arxiv.org/abs/1303.5076}{{\ttfamily 1303.5076}}].

\bibitem{wmap7}
E.~{Komatsu}, K.~M. {Smith}, J.~{Dunkley}, C.~L. {Bennett}, B.~{Gold},
  G.~{Hinshaw} et~al., \emph{{Seven-year Wilkinson Microwave Anisotropy Probe
  (WMAP) Observations: Cosmological Interpretation}},
  \href{https://doi.org/10.1088/0067-0049/192/2/18}{\emph{\apjs} {\bfseries
  192} (2011) 18} [\href{https://arxiv.org/abs/1001.4538}{{\ttfamily
  1001.4538}}].

\bibitem{bogdanovic2008}
T.~{Bogdanovi{\'c}}, B.~D. {Smith}, S.~{Sigurdsson} and M.~{Eracleous},
  \emph{{Modeling of Emission Signatures of Massive Black Hole Binaries. I.
  Methods}}, \href{https://doi.org/10.1086/521828}{\emph{\apjs} {\bfseries 174}
  (2008) 455} [\href{https://arxiv.org/abs/0708.0414}{{\ttfamily 0708.0414}}].

\bibitem{rossi2010}
E.~M. {Rossi}, G.~{Lodato}, P.~J. {Armitage}, J.~E. {Pringle} and A.~R. {King},
  \emph{{Black hole mergers: the first light}},
  \href{https://doi.org/10.1111/j.1365-2966.2009.15802.x}{\emph{\mnras}
  {\bfseries 401} (2010) 2021}
  [\href{https://arxiv.org/abs/0910.0002}{{\ttfamily 0910.0002}}].

\bibitem{corrales2010}
L.~R. {Corrales}, Z.~{Haiman} and A.~{MacFadyen}, \emph{{Hydrodynamical
  response of a circumbinary gas disc to black hole recoil and mass loss}},
  \href{https://doi.org/10.1111/j.1365-2966.2010.16324.x}{\emph{\mnras}
  {\bfseries 404} (2010) 947}
  [\href{https://arxiv.org/abs/0910.0014}{{\ttfamily 0910.0014}}].

\bibitem{kocsis2008}
B.~{Kocsis}, Z.~{Haiman} and K.~{Menou}, \emph{{Premerger Localization of
  Gravitational Wave Standard Sirens with LISA: Triggered Search for an
  Electromagnetic Counterpart}},
  \href{https://doi.org/10.1086/590230}{\emph{\apj} {\bfseries 684} (2008) 870}
  [\href{https://arxiv.org/abs/0712.1144}{{\ttfamily 0712.1144}}].

\bibitem{shields2003}
G.~A. {Shields}, K.~{Gebhardt}, S.~{Salviander}, B.~J. {Wills}, B.~{Xie}, M.~S.
  {Brotherton} et~al., \emph{{The Black Hole-Bulge Relationship in Quasars}},
  \href{https://doi.org/10.1086/345348}{\emph{\apj} {\bfseries 583} (2003) 124}
  [\href{https://arxiv.org/abs/astro-ph/0210050}{{\ttfamily
  astro-ph/0210050}}].

\bibitem{wetzel2010}
A.~R. {Wetzel} and M.~{White}, \emph{{What determines satellite galaxy
  disruption?}},
  \href{https://doi.org/10.1111/j.1365-2966.2009.16191.x}{\emph{\mnras}
  {\bfseries 403} (2010) 1072}
  [\href{https://arxiv.org/abs/0907.0702}{{\ttfamily 0907.0702}}].

\bibitem{garrison2017}
S.~{Garrison-Kimmel}, A.~{Wetzel}, J.~S. {Bullock}, P.~F. {Hopkins},
  M.~{Boylan-Kolchin}, C.-A. {Faucher-Gigu{\`e}re} et~al., \emph{{Not so lumpy
  after all: modelling the depletion of dark matter subhaloes by Milky Way-like
  galaxies}}, \href{https://doi.org/10.1093/mnras/stx1710}{\emph{\mnras}
  {\bfseries 471} (2017) 1709}
  [\href{https://arxiv.org/abs/1701.03792}{{\ttfamily 1701.03792}}].

\bibitem{blecha2008}
L.~{Blecha} and A.~{Loeb}, \emph{{Effects of gravitational-wave recoil on the
  dynamics and growth of supermassive black holes}},
  \href{https://doi.org/10.1111/j.1365-2966.2008.13790.x}{\emph{\mnras}
  {\bfseries 390} (2008) 1311}
  [\href{https://arxiv.org/abs/0805.1420}{{\ttfamily 0805.1420}}].

\bibitem{hoffman2007}
L.~{Hoffman} and A.~{Loeb}, \emph{{Dynamics of triple black hole systems in
  hierarchically merging massive galaxies}},
  \href{https://doi.org/10.1111/j.1365-2966.2007.11694.x}{\emph{\mnras}
  {\bfseries 377} (2007) 957}
  [\href{https://arxiv.org/abs/astro-ph/0612517}{{\ttfamily
  astro-ph/0612517}}].

\bibitem{sesana2004}
A.~{Sesana}, F.~{Haardt}, P.~{Madau} and M.~{Volonteri}, \emph{{Low-Frequency
  Gravitational Radiation from Coalescing Massive Black Hole Binaries in
  Hierarchical Cosmologies}}, \href{https://doi.org/10.1086/422185}{\emph{\apj}
  {\bfseries 611} (2004) 623}
  [\href{https://arxiv.org/abs/astro-ph/0401543}{{\ttfamily
  astro-ph/0401543}}].

\bibitem{amaro2010}
P.~{Amaro-Seoane}, A.~{Sesana}, L.~{Hoffman}, M.~{Benacquista}, C.~{Eichhorn},
  J.~{Makino} et~al., \emph{{Triplets of supermassive black holes:
  astrophysics, gravitational waves and detection}},
  \href{https://doi.org/10.1111/j.1365-2966.2009.16104.x}{\emph{\mnras}
  {\bfseries 402} (2010) 2308}
  [\href{https://arxiv.org/abs/0910.1587}{{\ttfamily 0910.1587}}].

\bibitem{oleary2009}
R.~M. {O'Leary} and A.~{Loeb}, \emph{{Star clusters around recoiled black holes
  in the Milky Way halo}},
  \href{https://doi.org/10.1111/j.1365-2966.2009.14611.x}{\emph{\mnras}
  {\bfseries 395} (2009) 781}
  [\href{https://arxiv.org/abs/0809.4262}{{\ttfamily 0809.4262}}].

\end{thebibliography}\endgroup
\bibliographystyle{JHEP}

\end{document}